\documentclass[journal]{IEEEtran}

\ifCLASSINFOpdf
\else
   \usepackage[dvips]{graphicx}
\fi
\usepackage{url}
\usepackage{graphicx}

\usepackage{cite}
\usepackage{physics,amsmath}
\usepackage{amssymb}
\usepackage{algorithm}
\usepackage{algpseudocode}
\usepackage{booktabs}
\usepackage{multirow}

\makeatletter
\newcommand{\algmargin}{\the\ALG@thistlm}
\makeatother
\newlength{\whilewidth}
\algdef{SE}[parWHILE]{parWhile}{EndparWhile}[1]
  {\parbox[t]{\dimexpr\linewidth-\algmargin}{%
     \hangindent\whilewidth\strut\algorithmicwhile\ #1\ \algorithmicdo\strut}}{\algorithmicend\ \algorithmicwhile}%
\algnewcommand{\parState}[1]{\State%
  \parbox[t]{\dimexpr\linewidth-\algmargin}{\strut #1\strut}}

\begin{document}

\title{SAR Despeckling using a Denoising Diffusion Probabilistic Model}

\author{Malsha V. Perera, \IEEEmembership{Student Member, IEEE}, Nithin Gopalakrishnan Nair, \IEEEmembership{Student Member, IEEE}, Wele Gedara Chaminda Bandara, \IEEEmembership{Student Member, IEEE} and Vishal M. Patel, \IEEEmembership{Senior Member, IEEE}

\thanks{This work was supported by the NSF CAREER Award under
Grant 2045489. }
\thanks{M. V. Perera, N. G. Nair, W. G. C. Bandara, and V. M. Patel are with the Department of Electrical and Computer Engineering, Johns Hopkins University, Baltimore, MD, 21218. (e-mail: \{jperera4, ngopala2, wbandar1, and vpatel36\}@jhu.edu).}
}

\markboth{Journal of \LaTeX\ Class Files, Vol. 14, No. 8, August 2015}
{Shell \MakeLowercase{\textit{et al.}}: Bare Demo of IEEEtran.cls for IEEE Journals}
\maketitle

\begin{abstract}
Speckle is a multiplicative noise which affects all coherent imaging modalities including Synthetic Aperture Radar (SAR) images. The presence of speckle degrades the  image quality and adversely affects the performance of SAR image understanding applications such as automatic target recognition and change detection. Thus, SAR despeckling is an important problem in remote sensing. In this paper, we introduce SAR-DDPM, a denoising diffusion probabilistic model for SAR despeckling. The proposed method comprises of a Markov chain that transforms clean images to white Gaussian noise by repeatedly adding random noise.  The despeckled image is recovered by a reverse process which iteratively predicts the added noise using a noise predictor which is conditioned on the speckled image. In addition, we propose a new inference strategy based on cycle spinning to improve the despeckling performance. Our experiments on both synthetic and real SAR images demonstrate that the proposed method achieves significant improvements in both quantitative and qualitative results over the state-of-the-art despeckling methods. The code is available at: \url{https://github.com/malshaV/SAR_DDPM}
\end{abstract}


\IEEEpeerreviewmaketitle

\section{Introduction}

\IEEEPARstart{S}{ynthetic}  Aperture Radar (SAR) is a coherent imaging modality which is strongly invariant to different environmental conditions. Therefore, SAR is widely used in applications such as landscape classification, disaster monitoring, and surface change detection. Nonetheless, SAR images are often affected by speckle which is a signal-dependent, spatially correlated noise. While degrading the SAR images, the presence of speckle creates an adverse effect on downstream tasks. Hence, several methods for speckle removal have been proposed in the literature for enhancing the SAR images over the past decades.

For a given SAR intensity $Y$, the mathematical model of multiplicative speckle noise $N$ can be expressed as follows:
\begin{equation}
   Y = XN
\label{eq1},
\end{equation}
where $X$ is the speckle-free or the clean image.  Generally, it is assumed that $N$  follows a Gamma distribution with unit mean and variance of $1/L$, with the following probability density function:
\begin{equation}
    p(N) = \frac{1}{\Gamma(N)}L^LN^{L-1}e^{-LN},
\label{eq2}
\end{equation}
where $\Gamma(.)$ is the Gamma function and $L$ is the number of looks in multilook processing. 

The first attempts at despeckling employed spatial domain filtering-based approaches such as Lee filter \cite{lee_filter}, Kuan filter \cite{kuan_filter} and Gamma maximum a \textit{posteriori} (MAP) filter \cite{gamma_filter}. These approaches generally use the spatial correlation
of image to filter noise by employing a sliding window to compute the pixel value at the center of the window. However, these methods result in 
intense smoothing and most edges and texture details in the SAR images are lost. In 2009, Deledalle {\em et al.} proposed a probabilistic patch-based (PPB) \cite{ppb} filtering method which combines non-local means filtering and 3-D block matching approach (BM3D) \cite{BM3D}. Parrilli  {\em et al.} \cite{sar_bm3d} introduced SAR-BM3D for SAR despeckling by extending BM3D. Sparse dictionary-based despeckling methods have also been proposed in the literature \cite{despecking_SR}.  A detailed review of SAR despeckling approaches can be found in \cite{review1} and \cite{review2}.

More recently, deep learning algorithms have gained popularity and achieved state-of-the-art performance in wide range of computer vision and image processing tasks. Following this trend, several studies have attempted to apply deep networks for SAR despeckling. SAR-CNN \cite{SARCNN} is one of the earliest Convolutional Neural Network (CNN) based despeckling methods. This method transforms the multiplicative speckle noise to additive noise applying a homomorphic transformation to the SAR images. In order to train the network, SAR-CNN uses multi-temporal fusion to generate clean reference images. Wang {\em et al.} \cite{IDCNN} proposed ID-CNN which is trained on synthetically speckled optical images. ID-CNN uses a residual network architecture, which obtains despeckled image by dividing the input speckled image by the estimated speckle. ID-GAN proposed in \cite{IDGAN}, uses a Generative Adversarial Network (GAN) \cite{GAN} for despeckling which is trained using a combination of  Euclidean loss, Perceptual loss and Adversarial loss. In addition to training deep learning models using spatial loss functions, the Multi-Objective network (MONet) \cite{MONET} incorporates a loss function that computes Kullback–Leibler divergence between the the predicted and simulated speckle probability distribution.  Studies such as multiscale residual dense dual attention network (MRDDANet) \cite{MRDDANet} and SAR-CAM \cite{SARCAM} incorporate various attention modules in the network architecture for despeckling. In \cite{SAR_ON}, an overcomplete CNN architecture (SAR-ON) which focuses on learning low-level features by restricting the receptive field is used for SAR despeckling. Apart from CNN architectures, \cite{SAR_T} introduces a Transformer-based architecture (Trans-SAR) for despeckling of SAR images. 

Recently, Denoising Diffusion Probabilistic Models (DDPM) \cite{DDPM} have emerged as an alternative approach for generative modelling. Dhariwal and Nichol \cite{Diffusion_beat_GAN} successfully presented that DDPM can outperform the current state-of-the-art GAN-based generative models \cite{GAN} in image synthesis. DDPM is parameterized by a Markov chain that gradually adds noise to the data until the signal is destroyed. During inference, a sample belonging to the training distribution can be generated by starting with a randomly sampled Gaussian noise and iteratively applying a reverse diffusion step. DDPM is trained by optimizing the variational lower bound of the negative log-likelihood of the data, and it avoids the mode collapse often encountered by GANs. With the success in image synthesis tasks, DDPMs have been adopted into a variety of vision tasks such as super-resolution \cite{SRDiff}, in-painting \cite{repaint} and image  restoration \cite{image_restore}. To the best of our knowledge, SAR despeckling based on Denoising Diffusion Probabilistic Models has not been studied in the literature.  

\begin{figure}
\centerline{\includegraphics[width=0.9\columnwidth]{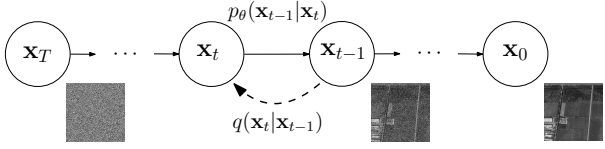}}
\caption{Overview of the forward and reverse diffusion process}
\label{DDPM process}
\end{figure}

Inspired by these studies, we propose SAR-DDPM, a Denoising Diffusion Probabilistic Model-based approach for SAR despeckling. In our approach, we iteratively recover the clean image by employing a noise predictor conditioned on the speckled image. We train the proposed DDPM model with synthetically speckled optical images and test our approach on both synthetic and real SAR images. During inference, we incorporate a novel strategy based on cycle spinning \cite{cycle_spin} to improve the despeckling performance. Finally, we demonstrate the effectiveness of our method on synthetic and real SAR images by comparing with several recent despeckling approaches. 


\vspace{-3mm}
\section{Proposed Method}

\begin{figure*}
\centerline{\includegraphics[width=\textwidth]{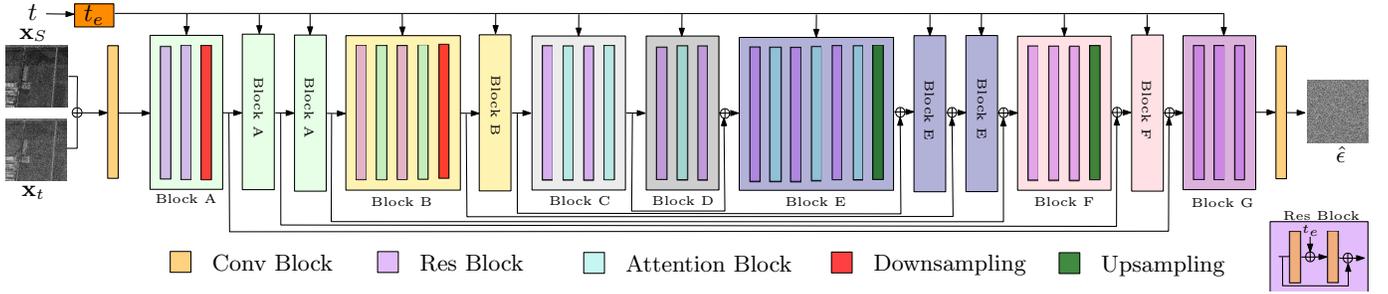}}
\caption{Overview of the conditional noise predictor network architecture in SAR-DDPM. Here, 2D-Convolution block, and residual block are denoted as ``Conv Block”, and ``Res Block”, respectively.}
\label{DDPM model}
\end{figure*}

\subsection{Denoising Diffusion Probabilistic Models}
Denoising Diffusion Probabilistic Models define a Markov chain that transforms an image $\vb{x}_0$ to white Gaussian Noise $\vb{x}_T \sim \mathcal{N}(0,1)$ by adding random noise in $T$ diffusion time steps. During inference, a random noise vector $\vb{x}_T$ is sampled and gradually denoised  until it reaches the desired image $\vb{x}_0$. As illustrated in Fig. \ref{DDPM process}, DDPM comprises of two processes: a forward diffusion process and a reverse diffusion process.

In the forward diffusion process, $\vb{x}_0 \sim q(\vb{x})$ sampled from the real data distribution is converted to $\vb{x}_T$ by gradually adding Gaussian noise $\epsilon$ according to a variance schedule $\beta_1, \cdots, \beta_T $
\begin{equation}
    q(\vb{x}_{t}|\vb{x}_{t-1}) = \mathcal{N}(\vb{x}_t; \sqrt{1-\beta_t}\vb{x}_{t-1}, \beta_T \vb{I}),\\
\end{equation} 
\begin{equation}
        q(\vb{x}_{1:T}|\vb{x}_0) = \prod_{t=1}^{T} q(\vb{x}_{t}|\vb{x}_{t-1}).
\end{equation}
In the forward process, $\vb{x}_t$ can be sampled at any arbitrary time step $t$ by setting $\alpha_t = 1- \beta_t$ and $\Bar{\alpha_t} = \prod_{t=1}^{T} \alpha_i$
\begin{equation}
    q(\vb{x}_{t}|\vb{x}_{0}) = \mathcal{N}(\vb{x}_t; \Bar{\alpha_t}\vb{x}_0, (1- \Bar{\alpha_t})\vb{I}).
\label{eq:q_sample}
\end{equation} 
This can be further reparameterized as follows
\begin{equation}
    \vb{x_t} = \sqrt{\Bar{\alpha_t}}\vb{x}_0 + \sqrt{1-\Bar{\alpha_t}}\epsilon, \; \epsilon \sim \mathcal{N}(0,1).
\end{equation}
The reverse diffusion process is modeled by a neural network trained to predict the parameters $\mu_\theta (\vb{x}_t,t)$ and $\Sigma_\theta(\vb{x}_t,t)$ of a Gaussian distribution 
\begin{equation}
    p_\theta(\vb{x}_{t-1}|\vb{x}_{t}) = \mathcal{N}(\vb{x}_{t-1}; \mu_\theta (\vb{x}_t,t), \Sigma_\theta(\vb{x}_t,t)).
    \label{p eq}
\end{equation} 
As reported by Ho {\em et al.} \cite{DDPM},  learning objective for the above model is derived by considering the variational lower bound,
\begin{multline}
   \mathbb{E}[p_\theta(\vb{x}_0)] \leq L = \mathbb{E}_q \bigg[\underbrace{D_{KL}(q(\vb{x}_T|\vb{x}_0)\,||\,p(\vb{x}_T))}_{L_T}\\
    + \sum_{t>1} \underbrace{D_{KL}(q(\vb{x}_{t-1}|\vb{x}_t, \vb{x}_0)\,||\,p(\vb{x}_{t-1}|\vb{x}_t))}_{L_{t-1}} \underbrace{- \text{log}\,p_\theta(\vb{x}_0|\vb{x}_1)}_{L_0} \bigg].
\label{KL_eq}
\end{multline} 
Note that the term $L_{t-1}$ is used to train the neural network and this can be computed in closed form as $L_{t-1}$ compares two Gaussian distributions. As $\vb{x}_t$ is available as an input during  training, the predicted mean $\mu_\theta (\vb{x}_t,t)$ can be reparameterized as follows,
\begin{equation}
    \mu_\theta (\vb{x}_t,t) = \frac{1}{\sqrt{\alpha_t}}\bigg( \vb{x}_t - \frac{\beta_t}{\sqrt{1-\Bar{\alpha}_t}}\epsilon_\theta(\vb{x}_t,t)\bigg).
    \label{mu eq}
    \end{equation}
For simplicity, Ho {\em et al.} \cite{DDPM} derive the following training objective from $L_{t-1}$ in Eq. \eqref{KL_eq}.
\begin{equation}
    L_{simple} = \mathbb{E}_{t,\vb{x}_0,\epsilon} \big[ \|\epsilon - \epsilon_\theta(\vb{x}_t,t)\|^{2} \big].
\end{equation}

\vspace{-2mm}
\subsection{SAR-DDPM}
The proposed method is based on a $T$ step DDPM model with a forward and a reverse diffusion processes as shown in  Fig. \ref{DDPM process}. During the forward diffusion process, we use the clean image $\vb{x}_C$ as the input image $\vb{x}_0$ which is converted into $\vb{x}_T$ by gradually adding a Gaussian noise. The reverse process uses the conditional noise predictor $\epsilon_\theta$ to recover the clean image $\vb{x}_C$ from $\vb{x}_T$ by denoising iteratively in $T$ steps. 

In the proposed method, the conditional noise predictor $\epsilon_\theta$ is trained to predict the noise added in each diffusion step conditioned on the speckled image $\vb{x}_S$. The network architecture of the proposed conditional noise predictor is illustrated in Fig. \ref{DDPM model}. The proposed network follows a U-Net \cite{UNet} like architecture which comprises of convolutional residual blocks, attention blocks, downsampling convolutions, upsampling convolutions, and skip connections connecting the contracting and expansive pathways. First, the speckled image $\vb{x}_S$ and $\vb{x}_t$ are concatenated and passed through a convolutional layer, while the timestep $t$ is transformed to the timestep embedding $t_e$ using the transformer sinusoidal positional encoding \cite{Attention}. The time step embedding $t_e$ is given as an input to each residual block as depicted in Fig. \ref{DDPM model}. Similar to \cite{Diffusion_beat_GAN}, we use self-attention blocks at multiple resolutions, and BigGAN \cite{BigGAN} residual blocks for upsampling and downsampling. Finally, the output from the last residual block of the expansive pathway is passed through a convolutional block to predict the noise $\hat{\epsilon}$ at the time step $t-1$. The predicted noise is then used to obtain $\vb{x}_{t-1}$ using Eq. \eqref{p eq} and \eqref{mu eq}.

\begin{figure}
\centerline{\includegraphics[width=0.85\columnwidth]{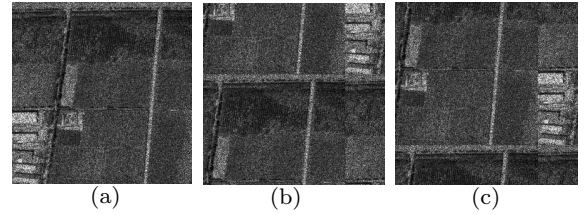}}
\caption{Cyclically spinned images with : (a) $u = 0$,$v= 0$ (Original speckled image) , (b) $u = 100$, $v= 200$, and (c) $u = 200$, $v= 200$ }
\label{Cycle spin}
\vspace{-3mm}
\end{figure}

Algorithm \ref{train algo} summarises the training procedure of the proposed method where synthetically speckled images and their corresponding clean images are used to train the SAR-DDPM in $T$ diffusion steps. During inference, we adopt the idea of cycle spinning \cite{cycle_spin} to improve the performance of SAR-DDPM. The function $f_{cs}(.,u,v)$ is used to shift an image cyclically by $u$ rows and $v$ columns as depicted in Fig. \ref{Cycle spin}. We despeckle the cyclically shifted images using the DDPM model, apply inverse cyclic shift, and average them to obtain the final despeckled image. The complete inference procedure is given in Algorithm \ref{test algo}.

\begin{algorithm}
\caption{Training}
\begin{algorithmic}[1]
\renewcommand{\algorithmicrequire}{\textbf{Input:}}
\renewcommand{\algorithmicensure}{\textbf{Initialize:}}
\Require Speckled image and clean image pairs $P = \{(\vb{x}_S^k,\vb{x}_C^k)\}^{K}_{k=1}$
\Repeat
    \State $(\vb{x}_S,\vb{x_C})\sim P$
    \State $t \sim \text{Uniform}(\{1,\ldots,T\})$
    \State $\epsilon\sim \mathcal{N}(\vb{0}, \vb{I})$
    \parState{%
          Take gradient descent step on $\nabla_\theta \,||\epsilon - \epsilon_\theta(\vb{x}_t,\vb{x}_S, t)||^{2}$,  $\vb{x}_t = \sqrt{\bar{\alpha_t}}\vb{x}_c + \sqrt{1-\bar{\alpha_t}\epsilon}$}

\Until {converged}
\end{algorithmic} 
\vspace{-1mm}
\label{train algo}
\end{algorithm}

\begin{algorithm}
\caption{Inference}
\begin{algorithmic}[1]
\renewcommand{\algorithmicrequire}{\textbf{Input:}}
\renewcommand{\algorithmicensure}{\textbf{Initialize:}}
\Require Speckled image $\vb{x}_S$, $\{u_i\}^{M}_{i=1}$ , $\{v_i\}^{M}_{i=1}$
\For {$i = 1, \ldots, M$}
\State $\vb{x}_i = f_{cs}(\vb{x}_S,u_i,v_i)$
\State Sample $\vb{x}_T\sim \mathcal{N}(\vb{0}, \vb{I})$
\For{$t = T, \ldots, 1$}
    \State sample $\vb{z}\sim \mathcal{N}(\vb{0},\vb{I})$ if $t>1$ else $\vb{z}=0$
    \parState{ %
    compute $\vb{x}_{t-1} = \frac{1}{\sqrt{\alpha_t}}\Big(\vb{x}_t - \frac{1-\alpha_t}{\sqrt{1-\bar{\alpha_t}}} \epsilon_\theta(\vb{x}_t,\vb{x}_i,t)\Big) + \sigma_t\vb{z}$}
\EndFor
\State $\vb{x}_{C_i}=f_{cs}^{-1}(\vb{x}_0,u_i,v_i)$
\State $\vb{x}_{sum} = \vb{x}_{sum} + \vb{x}_{C_i}$ if $i>1$ else $\vb{x}_{sum} = \vb{x}_{C_i}$
\EndFor\\
\Return $\vb{x}_{C} = \frac{\vb{x}_{sum} }{M}$
\end{algorithmic}
\label{test algo}
\end{algorithm}

\begin{figure*}
\centerline{\includegraphics[width=\textwidth]{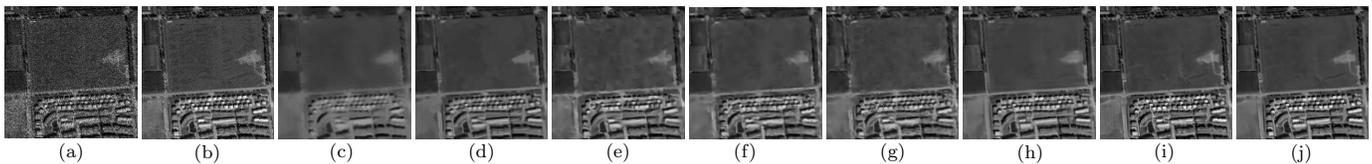}}
\caption{Results on a synthetically speckled image: (a) Speckled image, (b) Ground Truth, (c) PPB, (d) SAR-BM3D, (e) ID-CNN, (f) Trans-SAR, (g) SAR-ON, (h) SAR-CAM, (i) SAR-DDPM, (j) SAR-DDPM + Cycle Spinning .}
\label{synthetic image}
\vspace{-2mm}
\end{figure*}

\begin{figure*}
\centerline{\includegraphics[width=\textwidth]{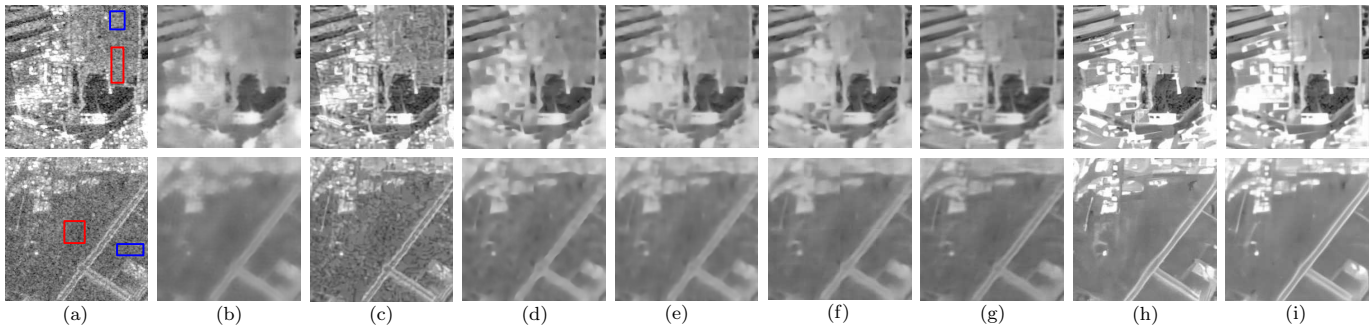}}
\caption{Results on real SAR images: (a) SAR image, (b) PPB, (c) SAR-BM3D, (d) ID-CNN, (e) Trans-SAR, (f) SAR-ON, (g) SAR-CAM, (h) SAR-DDPM, (i) SAR-DDPM + Cycle Spinning .}
\label{SAR image}
\vspace{-2mm}
\end{figure*}


\section{Experiments and Results}
In this section, we present the experiments and results of our proposed method on both synthetically speckled images and real SAR images. We compare the performance of SAR-DDPM with the following despeckling algorithms: PPB \cite{ppb}, SAR-BM3D \cite{sar_bm3d}, ID-CNN \cite{IDCNN}, Trans-SAR \cite{SAR_T}, SAR-ON \cite{SAR_ON}, and SAR-CAM \cite{SARCAM}. Following Eq. \eqref{eq1} and \eqref{eq2} , we created single-look ($L=1$) synthetic speckle images using 15k remote sensing images from publicly available DSIFN \cite{DSIFN} dataset in order to train SAR-DDPM. The network was trained on an NVIDIA RTX 2080Ti GPU for 30k iterations and the network weights were initialized with pretrained ImageNet weights from \cite{Diffusion_beat_GAN}. In this work, we set $T = 1000$ and $\{u_i\} = \{v_i\} = \{0,100,200\}$.

Table \ref{results_synthetic} shows the performance of the proposed algorithm in terms of average Peak Signal-to-Noise Ratio (PSNR) and Structured Similarity Index (SSIM) on 10 synthetically speckled images from the DSIFN dataset. Note that ID-CNN, Trans-SAR, SAR-ON, and SAR-CAM were also trained on the same synthetic data as the proposed method. As can be seen from Table \ref{results_synthetic}, SAR-DDPM significantly outperforms both classical and deep learning-based despeckling methods in terms of PSNR and SSIM. Moreover, by the results in Table \ref{results_synthetic} and Fig. \ref{synthetic image}, we can observe that cycle spinning significantly improves the despeckling performance of SAR-DDPM. Also, it can be seen from Fig. \ref{synthetic image} that our proposed method provides a better despeckling performance while preserving the fine structural details present in the speckled images compared to the other despeckling methods.

\begin{table}[h]
\setlength\tabcolsep{0pt}
\caption{Results on synthetically speckled images of DSIFN.} \label{results table}
\centering
\smallskip
\begin{tabular*}{\columnwidth}{@{\extracolsep{\fill}}ccc}
\toprule
Method & PSNR & SSIM  \\
 \midrule
  PPB \cite{ppb}& 23.96 dB & 0.62\\
  SAR-BM3D \cite{sar_bm3d}& 25.69 dB & 0.75 \\
  ID-CNN \cite{IDCNN}& 27.30 dB & 0.72\\
  Trans-SAR \cite{SAR_T} & 27.08 dB & 0.72 \\
  SAR-ON \cite{SAR_ON}& 27.16 dB & 0.73  \\
  SAR-CAM \cite{SARCAM}& 27.96 dB& 0.76 \\
  SAR-DDPM & 27.99 dB& 0.77 \\
  SAR-DDPM with cycle spinning & \textbf{29.42 dB} & \textbf{0.81}\\
\bottomrule
\label{results_synthetic}
\end{tabular*}
\vspace{-2mm}
\end{table}

In order to further evaluate the despeckling ability of our proposed method, the above despeckling methods are tested on 2 single-look SAR images of size $256 \times 256$ acquired using Sentinel-I \cite{sentinel}. The Equivalent Number of Looks (ENL) is an index suitable for evaluating the level of smoothing in homogeneous areas of SAR images when the clean ground truth images are unavailable \cite{SAR_tutorial}. ENL is defined as the ratio between the square of the mean and the variance of a homogeneous region. Table \ref{results real table} summarizes the quantitative comparison of despeckling results of SAR images in terms of ENL. The regions selected for the ENL value calculation are marked in red and blue boxes in Fig. \ref{SAR image}.  The proposed method resulted in the highest ENL values in all 4 regions which signifies the best despeckling performance out of the considered despeckling methods. From Fig. \ref{SAR image}, we can observe that quantitative results using ENL are consistent with the visual results.

In Fig. \ref{SAR image}, the PPB despeckling approach over-smooths the SAR image, destroying edges and structural details. It can also be observed that, SAR-BM3D preserves significantly more details than PPB. However, as indicated by the ENL values, SAR-BM3D exhibits a poor performance when it comes to smoothing. ID-CNN, Trans-SAR, and SAR-ON preserve more textural details than PPB and remove more speckle than SAR-BM3D. Compared to ID-CNN, Trans-SAR and SAR-ON, SAR-CAM retains more textural details while minimizing distortions in homogeneous regions. Even though SAR-CAM showcase a better despeckling ability than the previous methods, it can be observed that SAR-CAM blurs out some fine edges and structural details. While SAR-DDPM is able to recover these finer details than SAR-CAM, it creates slight distortions in the homogeneous areas resulting in lower ENL values. As evident from Fig. \ref{SAR image} (i), the use of cycle spinning along with SAR-DDPM reduces these distortions and recovers finer structural details than SAR-CAM.


\begin{table}[h]
\setlength\tabcolsep{0pt}
\caption{Estimated ENL values on real SAR images} 
\label{results real table}
\centering
\smallskip
\vspace{-2mm}
\begin{tabular*}{\columnwidth}{@{\extracolsep{\fill}}ccccc}
\toprule
Method & Image 1 & Image 1 & Image 2 &  Image 2 \\
& Red & Blue & Red & Blue \\
 \midrule
  PPB \cite{ppb}& \text{659.72} & \text{622.61} & \text{257.50} & \text{441.37}\\
  SAR-BM3D \cite{sar_bm3d}& \text{100.79} & \text{78.87} & \text{70.95} & \text{80.93} \\
  ID-CNN \cite{IDCNN}& \text{501.43} & \text{394.91} & \text{210.65} & \text{400.88}\\
  Trans-SAR \cite{SAR_T} & \text{613.61} & \text{485.48} & \text{234.69} & \text{406.54} \\
  SAR-ON \cite{SAR_ON} & \text{494.63} & \text{527.85} & \text{255.50} & \text{454.36}\ \\
  SAR-CAM \cite{SARCAM}& \text{566.20} & \text{628.49} & \text{271.88} & \text{500.11} \\
  SAR-DDPM & \text{446.69} & \text{301.84} & \text{256.39} & \text{190.99} \\
  SAR-DDPM with cycle spinning &\textbf{747.92} & \textbf{988.85} & \textbf{372.93} & \textbf{560.61}\\
\bottomrule
\end{tabular*}
\end{table}

\section{Conclusion}

We proposed SAR-DDPM, the first diffusion-based model for SAR despeckling. Our work performs despeckling through a novel strategy based upon ensembling multiple cycle spinning based estimates for the despeckled image generated using the diffusion model. Each estimate is obtained by starting from random Gaussian noise and iteratively denoising the latent variables using a noise predictor conditioned on the corresponding transformed speckled images. Our experiments on synthetic and real SAR images show promising quantitative and qualitative improvements compared with several existing despeckling methods. The proposed method proved to be effective in removing speckle while retaining fine structural details in the SAR images.

\bibliographystyle{IEEEtran}
\bibliography{refs}




\end{document}